\documentclass[preprint,12pt]{elsarticle}

\usepackage{amssymb}

\usepackage{longtable}
\usepackage[reviewcopy]{adndt}

\newcommand{\araa}{{Ann.\ Rev.\ Astron.\ Astrophys.}}		
\newcommand{\apj}{{ApJ}}

\journal{Atomic Data and Nuclear Data Tables}

\begin{document}

\begin{frontmatter}



\title{Sensitivity study for $s$ process nucleosynthesis in AGB stars}  

\author[GUF,GSI,Nugrid]{A.~Koloczek}
\address[GUF]{Goethe Universit{\"a}t, Frankfurt a.M., 60438, Germany}
\address[GSI]{GSI Helmholtzzentrum f\"ur Schwerionenforschung GmbH, Darmstadt, 64291, Germany}
\author[GUF,Nugrid]{B.~Thomas}
\author[GUF,GSI]{J.~Glorius}
\author[GUF,GSI]{R.~Plag}
\author[UBas,Nugrid]{M.~Pignatari}
\author[GUF,Nugrid]{R.~Reifarth}
\ead{reifarth@physik.uni-frankfurt.de}
\author[GUF,UVic,Nugrid]{C.~Ritter}
\author[GUF]{S.~Schmidt}
\author[GUF]{K.~Sonnabend}

\address[UBas]{Department of Physics, University of Basel, Klingelbergstrasse 82, CH-4056 Basel, Switzerland}
\address[UVic]{University of Victoria, FP.O. Bos 3055, Victoria, B.C., V8W 3P6, Canada}
\address[Nugrid]{NuGrid collaboration, http://www.nugridstars.org}

\begin{abstract}

In this paper we present a large-scale sensitivity study of reaction rates in the main component of the $s$ process. The aim of this study is to identify all rates, which have a global effect on the $s$ process abundance distribution and the three most important rates for the production of each isotope. We have performed a sensitivity study on the radiative $^{13}$C-pocket and on the convective thermal pulse, sites of the $s$ process in AGB stars. We identified 22 rates, which have the highest impact on the $s$-process abundances in AGB stars. 



\end{abstract}

\end{frontmatter}

\section{Introduction \label{intro}}

In the solar system about half of the elements heavier than
iron are produced by the slow neutron capture process, or $s$ process
\cite{AKW99b}. The $s$ process is a sequence of neutron capture reactions on stable nuclei until an unstable isotope 
is produced, which usually decays via a $\beta^-$ decay to the element with the next higher proton number. This chain of neutron captures and beta decays will continue along the valley of stability up to $^{209}$Bi \cite{KGB11}.
The signature of the $s$ process contribution 
to the solar abundances
suggests a main, a weak and a strong component. While the main component is responsible for the 
atomic mass 
region from 90 to 209, the weak component contributes to the mass region
between 60 and 90. Finally, the strong component is required for the production of lead. The main and strong component is made by low mass 
stars with $1 \leq M/M_{\odot} \leq 3$ at different metallicities, whereas the weak component
is related to massive stars with $M \geq 8M_{\odot}$ ($M_{\odot}$ 
stands for the solar mass) \cite{PGH10}. 
According to our current understanding of the main $s$ process component, 
two alternating stellar burnings create environments with 
neutron densities of 10$^{6-7}$ cm$^{-3}$ and 10$^{11-12}$ cm$^{-3}$. The corresponding
neutron sources are the $^{13}$C($\alpha$,n)$^{16}$O and the $^{22}$Ne($\alpha$,n)$^{25}$Mg reaction.
 These reactions are activated in low-mass Asymptotic Giant Branch stars (AGB stars) \cite{LHL03}. AGB stars are characterized by alternating hydrogen shell burning and helium shell burning after the formation of a degenerate carbon-oxygen core. 

In this paper, we provide a complete sensitivity study for the final, 
most important pulse and the preceding $^{13}$C-pocket computed for the stellar model of a $3M_{\odot}$ star with metallicity $Z=0.02$. 


\section{$s$-process}

The production site for the main $s$ process component is located in thermally pulsing AGB stars, which is an advanced burning phase of low mass stars, where the core consists of degenerate oxygen and carbon and the helium inter-shell and the hydrogen envelope burn alternately.

During the AGB evolution phase, the $s$ process is mainly activated in the radiative $^{13}$C-pocket by the $^{13}$C($\alpha$,n)$^{16}$O reaction. After a thermal pulse (TP, \cite{ScH65}), the shell H burning is not efficient and H-rich material from the envelope is mixed down in the He intershell region by the so called Third Dredge Up (TDU, \cite{IbR83}). Convective boundary mixing (CBM) processes leave a decreasing abundance profile of protons below the bottom of the TDU. Protons are then captured by the He burning product $^{12}$C and converted to $^{13}$C via the channel $^{12}$C(p,$\gamma$)$^{13}$N($\beta^+$)$^{13}$C. Therefore, a $^{13}$C-rich radiative layer is formed, where the $^{13}$C($\alpha$,n)$^{16}$O reaction is activated before the occurrence of the next convective TP, at temperatures around $0.1$~GK and with neutron densities between 10$^{6}$ and 10$^{7}$~cm$^{-3}$. In particular, the $^{13}$C-pocket is the region where $^{13}$C is more abundant than the neutron poison $^{14}$N (for recent reviews, see \cite{Her05,SGC06}).

A smaller contribution to the $s$ process economy is given by the partial activation of the $^{22}$Ne($\alpha$,n)$^{25}$Mg reaction, during the convective TP.
The neutron source $^{22}$Ne produces only a few per cent of all the neutrons made by the $^{13}$C($\alpha$,n)$^{16}$O in the $^{13}$C-pocket, but it is activated at higher temperatures resulting in a higher neutron density (around 10$^{10}$ cm$^{-3}$). This affects the $s$-process abundance distribution for several isotopes along the $s$-process path (e.g. \cite{GAB98,LHL03}).
The most sensitive isotopes to the $^{22}$Ne($\alpha$,n)$^{25}$Mg contribution are located at the branch points.

\subsection{Branch points}

Branch points are unstable nuclei along the $s$-process path with a life time comparable to the neutron capture time. The average neutron capture time for the $s$ process depends on the isotope's (n,$\gamma$) cross section and the neutron density. It is around 10 years during the $^{13}$C phase. If the $s$-process path reaches such a nucleus, the path will split into two branches, with some of the mass flow following the $\beta$ decay and the rest of the mass flow following the neutron capture branch. The branching itself is very sensitive to the neutron capture time, hence the neutron density and the (n,$\gamma$) cross section. With increased neutron density, the neutron capture will become more likely and the beta decay less frequent and vice versa. 

\section{Nuclear Network}
\subsection{MACS}

For exact simulations it is essential to know the precise probability that a given reaction will take place.
 Taking into account the Maxwell-Boltzmann-distribution of the neutrons in stars, the cross sections can be 
 calculated by 
 \begin{equation}
\langle\sigma\rangle := \frac{\langle\sigma v\rangle}{v_T } = \frac{1}{v_T } \frac{\int \sigma v \Phi(v)\mathrm{d}\,v}{\int \Phi(v)\mathrm{d}\,v}
 \end{equation}
 where $\langle\sigma\rangle$ is the Maxwellian-averaged cross section (MACS). 
 $\langle\sigma v
\rangle$ is the integrated cross section $\sigma$ over the velocity distribution $\Phi(v)$ and 
\begin{equation}
v_T = (2kT/m)^{1/2}
\end{equation}
 with $m$ the reduced mass of the reaction partners. 

\subsection{Rates}

The reaction rate gives the change of abundance per unit time for one nucleus $X$ reacting with a particle $Y$. 
These rates, essential for the nucleosynthesis simulations, can be calculated by 
\begin{equation}
r=N_x N_y \langle\sigma v
\rangle (1+\delta_{xy})^{-1}
\end{equation}
where $N_x$ and $N_y$ is the number of nuclei $X$ and $Y$ per unit volume. 
The change of abundance per time is given by
\begin{equation}
(dN_x/dt)_y = -(1+\delta_{xy})~r
\end{equation}

Measuring exact values of the MACS and reaction rates can be quite difficult. There are still rates that have only been estimated theoretically. 

\subsection{Sensitivity studies}

Since some crucial rates (e.g. $^{85}$Kr(n,$\gamma$) \cite{RTR13}) along the $s$-process path are not known to sufficient precision, predictions based on rates have significant uncertainties \cite{RLK14}. In order to account for these uncertainties in isotopic abundances, it is essential to know the influence of these reactions on the resulting abundances. The sensitivity gives the coupling between the change in the rate and the change in the final abundance:

\begin{equation} 
 s_{ij} = \frac{\Delta N_j / N_j}{\Delta r_i / r_i}
\end{equation}

The sensitivity $s_{ij}$ is the ratio of the relative change in abundance $\Delta N_j / N_j$ of isotope $j$ and the relative change of the rate $\Delta r_i / r_i$.
In order to extract the sensitivity of a certain rate, simulations with a change in this rate are compared with the default run. A positive sensitivty means that an increase in the rate results in an increase of the final abundance, whereas a negative
sensitivity will decrease the final abundance with an increased rate. In this paper, we distinguish between global sensitivities, which affect the overall neutron density, and local sensitivities, which affect the $s$-process path in the vicinity of the nuclei under study.

\section{NuGrid}

The NuGrid collaboration provided a $3M_{\odot}$ and $Z=0.02$ stellar model and the tools to post-process this model for this study. The stellar model was calculated with the MESA (Modules for Experiments in Stellar Astrophysics) code and post-processed with MPPNP (Multizone Post Processing Network Parallel) the multi-zone driver of the PPN (Post Processing Network) code. 

For the sensitivity studies of the TP we recalculated in MPPNP all cycles of the last TP of the stellar model with changed nuclear network settings.

For the sensitivity studies of the radiative $^{13}$C-pocket we extracted a trajectory at the center of the $^{13}$C-pocket layers. Consistent initial abundances have been adopted for the simulations.

The network data in the PPN physics package is taken from a broad range of single rates and widely used reaction compilations. Focusing on charged-particle-induced reactions on stable isotopes in the mass range $A = 1$-$28$, the NACRE compilation \cite{AAR99} covers the main part of these reactions. Proton-capture rates from Iliadis et al. \cite{IDS01} in the mass range 20-40 are also included.
Neutron capture reaction rates are used from the KADoNiS project \cite{DHK06}, which combines the rates from earlier compilations of e.g. Bao et al. \cite{BBK00}. Beta-decay rates for unstable isotopes are taken from \cite{FFN85,OHM94,LaM00}. Further rates are taken from the Basel REACLIB compilation.
All reactions build up a reaction network of 14020 n-capture, charged particle and decay reactions. 
Within the MPPNP code a radial grid is used as the existing network is solved at each grid point. The size of the network is dynamically adapted depending on the conditions at each grid point. Calculations for mixing and nucleosynthesis are done with an implicit Newton-Raphson solver in operator split mode \cite{HDF08}. 

\section{Simulations}

With two single-zone trajectories at the bottom of the TP and the center of the $^{13}$C-pocket we checked for the importance of each rate by changing it in the network. Those showing an impact during the thermal pulse were recalculated with multiple zones and an increase and decrease of the rates  by 10$\%$. For the $^{13}$C-pocket all reactions were simulated with an increase and decrease of the rates by 5$\%$ and by 20$\%$. In this regime the sensitivity is constant, hence the sensitivities for changes of 10\% were averaged and tabulated.

Extensive simulations showed that the individual sensitivities of selected rates within each thermal pulse and $^{13}$C-pocket do not change significantly over the pulse history of the star. These results justify the assumption that the sensitivities extracted from a single event are representative for reoccurring events of the same type.

\section{Results}

\subsection{General sensitivity study}

In this part of our analysis we identified all rates with a global effect on the $s$-process abundances for the thermal pulse and the $^{13}$C-pocket and listed them in tables \ref{tab:TPglob} and \ref{tab:C13glob}. Furthermore, the three strongest rates that affect individual isotopes (averaged results for changes of $\pm$10\%) are listed in the online version of the full manuscript \cite{KTG16}. Only sensitivities greater than $\pm 0.1$ are reported there.

Those rates which have a global impact on the $s$-process abundances were differentiated into neutron donators, neutron poisons and competing captures.

\begin{figure}[ht]
   \includegraphics[width=\textwidth]{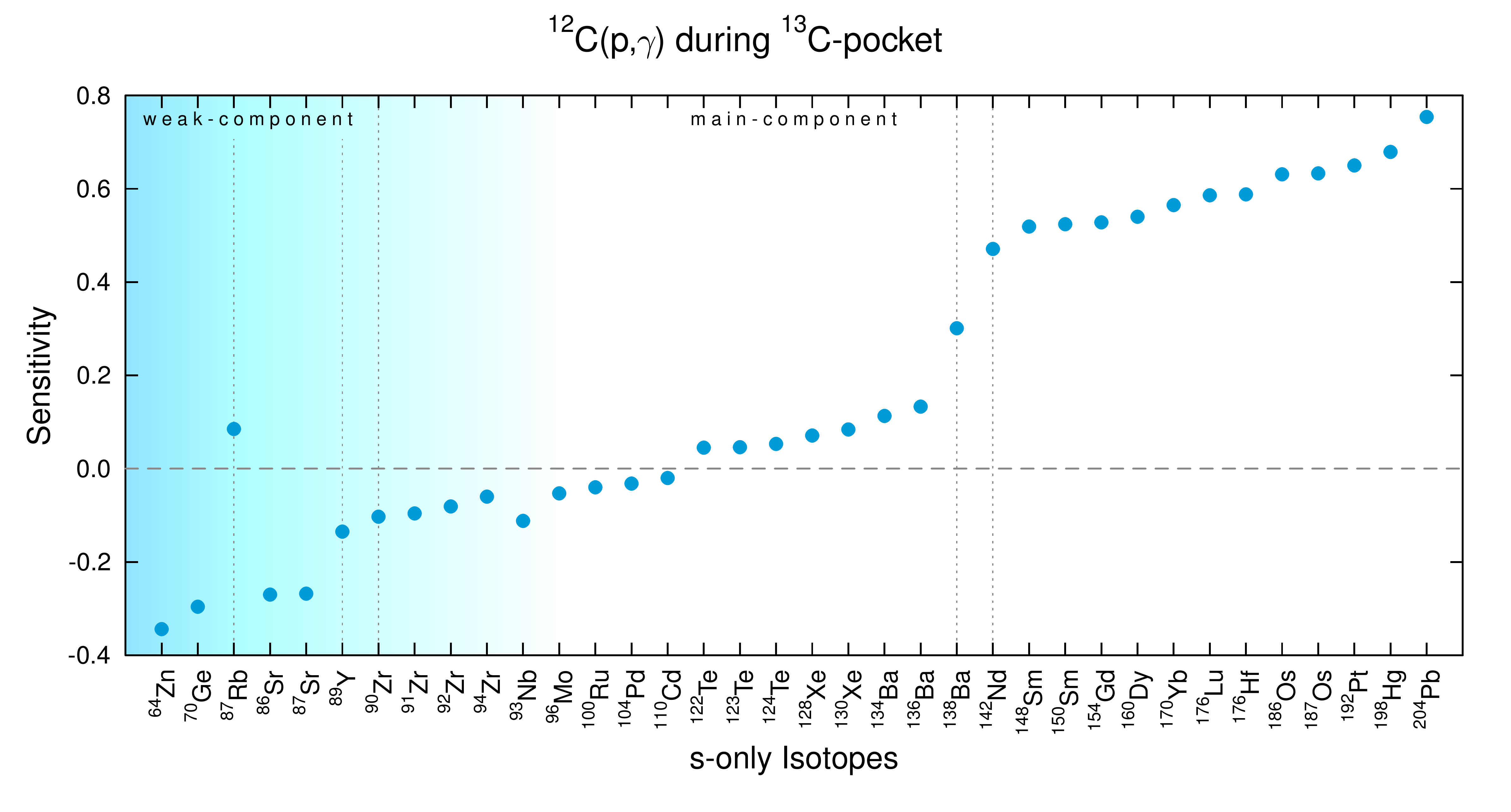}
   \caption{Sensitivity plot of the indirect neutron donator $^{12}$C(p,$\gamma$) in the $^{13}$C-pocket. The sensitivity is plotted over the $s$-only isotopes as well as $^{64}$Zn and $^{70}$Ge. The blue color gradient marks the weak $s$-process region. The vertical grey dotted lines are plotted on neutron magic isotopes. An increased neutron production leads to a higher production of heavy isotopes (mass number 110-210) and a stronger depletion of low mass isotopes (mass number 60-110). Neutron shell closures at $N = 50, 82$ are clearly visible as steps at $A \sim 90, 140$.\label{fig:donator}}
\end{figure}

Neutron donators are reactions, which either set neutrons free for the $s$ process or produce isotopes that ultimately set neutrons free. For example, the $^{13}$C($\alpha$,n) reaction is a direct neutron donator whereas the $^{12}$C(p,$\gamma$) reaction is an indirect neutron donator, since it leads to the production of the direct neutron donator $^{13}$C. A neutron donator is shown in figure \ref{fig:donator}.

\begin{figure}[ht]
   \includegraphics[width=\textwidth]{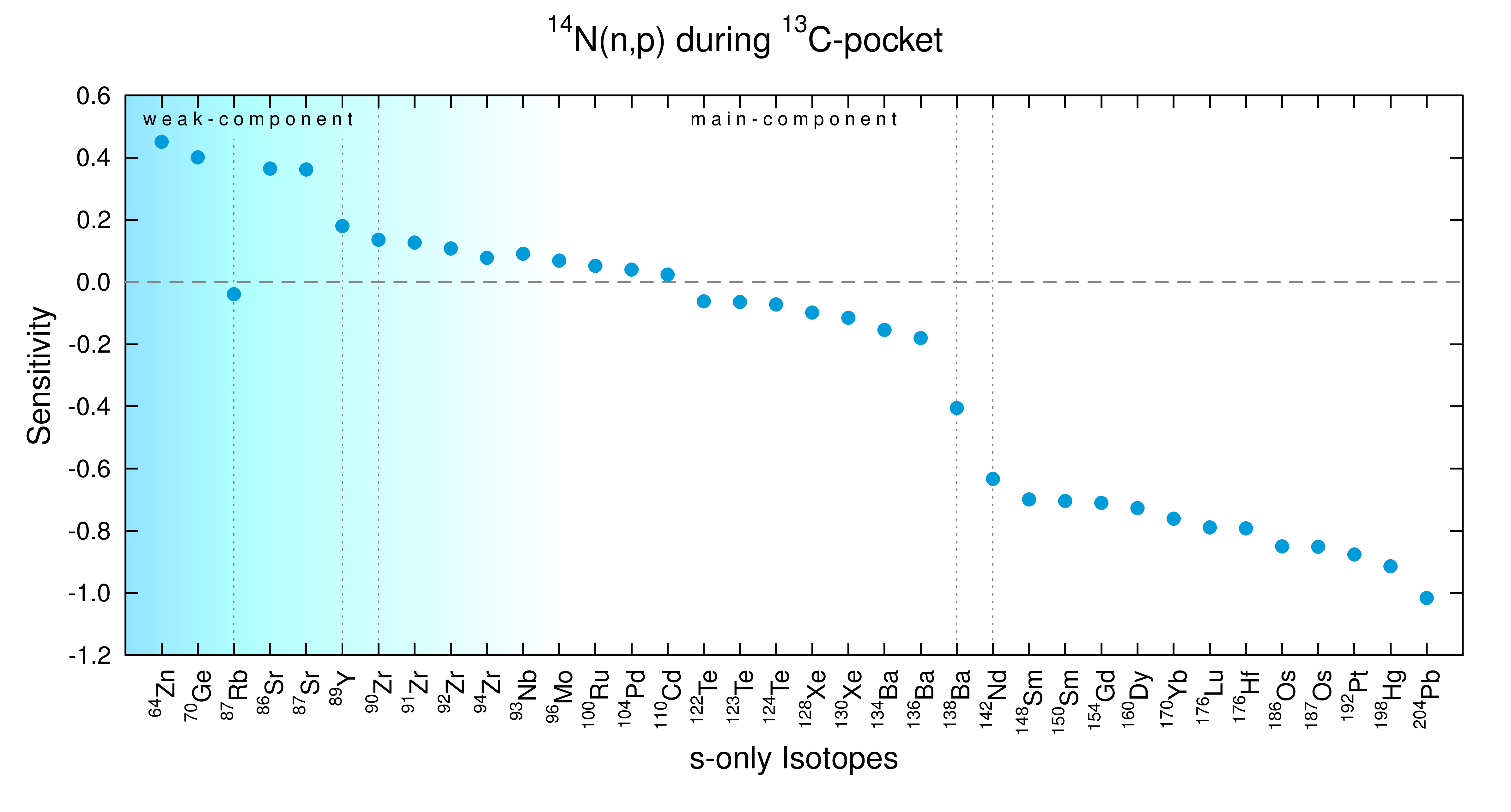}
   \caption{Sensitivity plot of the neutron poison $^{14}$N(n,p) in the $^{13}$C-pocket. An increased neutron capture of this neutron poison leads to a lower production of heavy isotopes (mass number 120-210) and a lower depletion of low mass isotopes (mass number 60-120).\label{fig:poison}}
\end{figure}

Neutron poisons are light isotopes with a sufficiently large neutron capture cross section to impact the neutron density or reactions, which produce these isotopes, or reactions, which compete with the neutron donator reactions. A neutron poison, which acts in all three ways, is, for example, the $^{14}$N(n,p) reaction, which not only consumes neutrons, but also produces protons, which will eventually compete with the $^{13}$C($\alpha$,n) reaction via the $^{13}$C(p,$\gamma$) reaction, which leads furthermore to the production of more $^{14}$N. Another example for a competing reaction acting as neutron poison is the $^{14}$C($\alpha$,$\gamma$) reaction as it requires $\alpha$ particles, which are crucial for the neutron source  $^{13}$C($\alpha$,n). A neutron poison is shown in figure \ref{fig:poison}.

\begin{figure}[ht]
   \includegraphics[width=\textwidth]{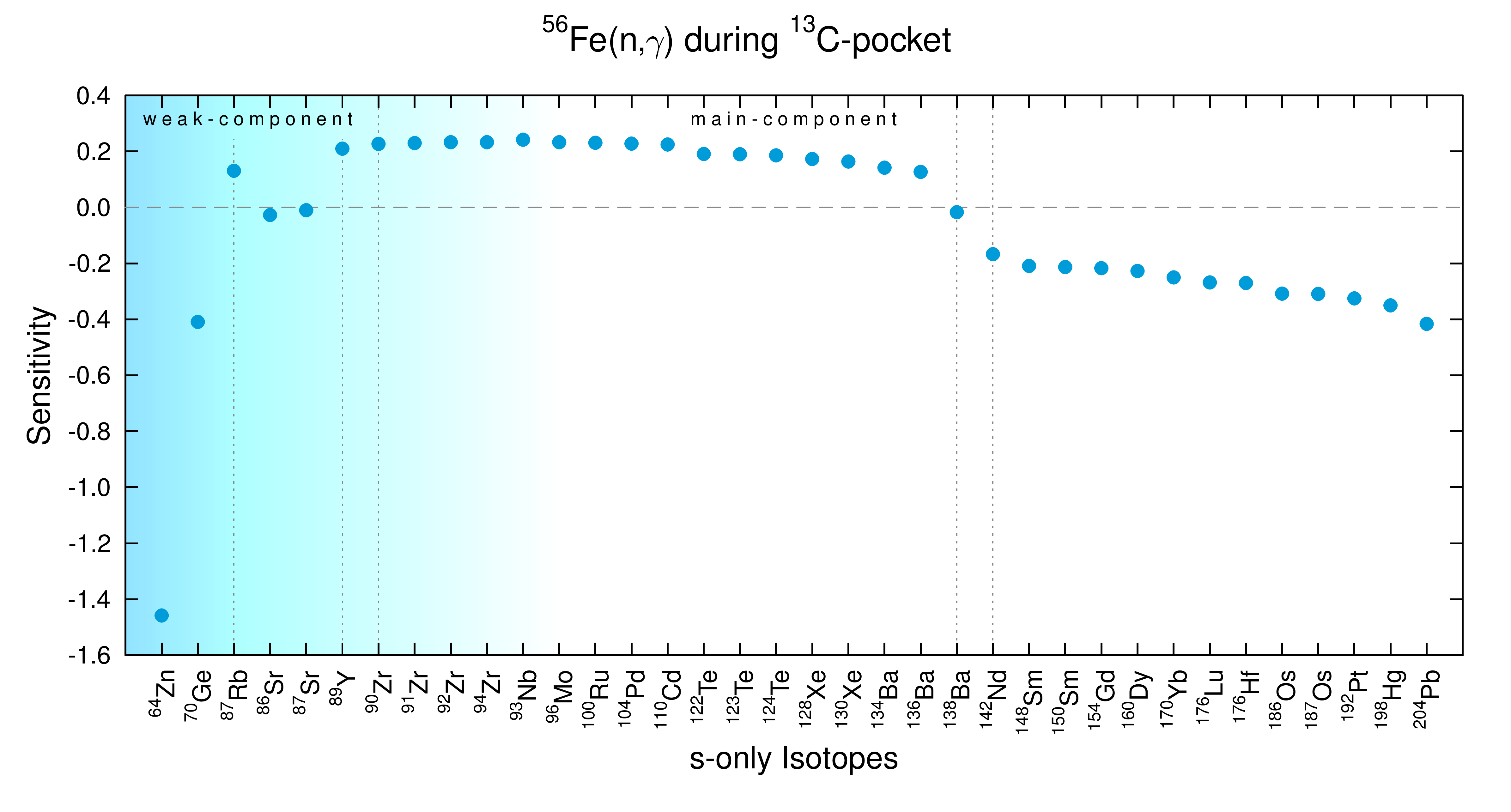}
   \caption{Sensitivity plot of the competing capture $^{56}$Fe(n,$\gamma$) in the $^{13}$C-pocket. An increased neutron capture of this isotope leads to a lower neutron per seed ratio. This lowers the production of isotopes in the mass regime of 140-210 and increases the abundance of isotopes in the mass regime 90-140.\label{fig:competing}}
\end{figure}

Competing captures are isotopes on the $s$-process path, which have a large neutron capture cross section or are abundant enough to affect the overall $s$-process evolution, which can be observed on many neutron magic isotopes. An example is the $^{56}$Fe(n,$\gamma$) reaction, which supports the $s$ process but impacts the amount of neutrons per seed, which shifts the peak in the production of isotopes from higher to lower mass numbers. A competing capture is demonstrated in figure \ref{fig:competing}.

\begin{table}[h]
\caption{Strongest globally affecting reactions during the TP, sorted by their impact. Only few rates have a global influence, because the TP has a short life-span and is convective. Cumulative effects will therefore not account under these conditions. The impact is given by the number of affected isotopes with a sensitivity over the threshold of $\pm 0.1$.\label{tab:TPglob}}
\begin{center}
\begin{tabular}{r|l|l}
 reaction & type of effect & affected isotopes\\
\hline
$^{22}$Ne($\alpha$,n) & neutron donator & 191\\
$^{25}$Mg(n,$\gamma$) & neutron poison  & 67\\
$^{142}$Nd(n,$\gamma$) & competing capture & 41\\
$^{144}$Nd(n,$\gamma$) & competing capture & 41\\
$^{56}$Fe(n,$\gamma$) & competing capture & 38\\
$^{140}$Ce(n,$\gamma$) & competing capture & 33\\
$^{146}$Nd(n,$\gamma$) & competing capture & 29\\
$^{22}$Ne(n,$\gamma$) & neutron poison & 25\\
$^{94}$Zr(n,$\gamma$) & competing capture & 24\\
$^{141}$Pr(n,$\gamma$) & competing capture & 23\\
$^{58}$Fe(n,$\gamma$) & competing capture & 21\\
\end{tabular}
\end{center}
\end{table}

\begin{table}[h]
\caption{Strongest globally affecting reactions during the $^{13}$C-pocket, sorted by their impact. The impact is given by the number of affected isotopes with a sensitivity over the threshold of $\pm 0.1$. \label{tab:C13glob}}
\begin{center}
\begin{tabular}{r|l|l}
 reaction & type of effect & affected isotopes\\
\hline
$^{56}$Fe(n,$\gamma$) & competing capture & 196\\
$^{64}$Ni(n,$\gamma$) & competing capture & 183\\
$^{14}$N(n,p) & neutron poison & 175 \\
$^{12}$C(p,$\gamma$) & neutron donator & 158\\
$^{13}$C(p,$\gamma$) & neutron poison & 150 \\
$^{16}$O(n,$\gamma$) & neutron poison & 145 \\
$^{22}$Ne(n,$\gamma$) & neutron poison & 144\\
$^{88}$Sr(n,$\gamma$) & competing capture & 131\\
$^{13}$C($\alpha$,n) & neutron donator & 114\\
$^{58}$Fe(n,$\gamma$) & competing capture & 112\\
$^{14}$C($\alpha$,$\gamma$) & neutron poison & 102 \\
$^{14}$C($\beta^\mathrm{-}$) & neutron poison & 95 \\
$^{138}$Ba(n,$\gamma$) & competing capture & 95\\
$^{140}$Ce(n,$\gamma$) & competing capture & 93\\
$^{139}$La(n,$\gamma$) & competing capture & 92\\
$^{142}$Nd(n,$\gamma$) & competing capture & 87\\
\end{tabular}
\end{center}
\end{table}

\begin{figure}[h]
   \includegraphics[width=\textwidth]{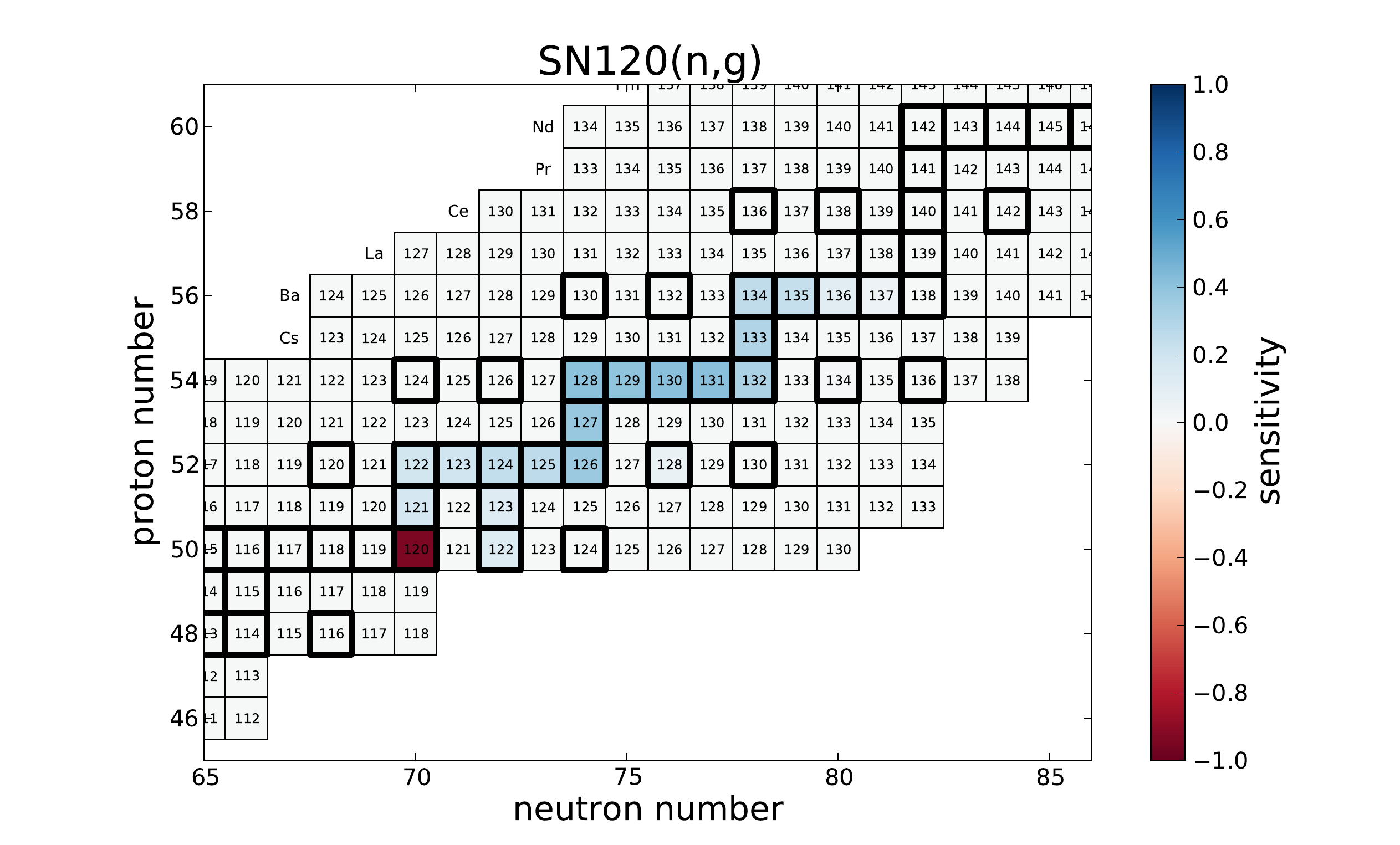}
   \caption{Sensitivity chart of the locally affecting $^{120}$Sn(n,$\gamma$) rate during the TP. An increased neutron capture rate of this isotope leads to a higher production of following isotopes on the $s$ process path.\label{fig:local_sens}}
\end{figure}

Local sensitivities are rates, which influence the production or depletion of isotopes in their neighborhood on the chart of nuclides. A locally sensitive rate is demonstrated in figure \ref{fig:local_sens}.

\subsection{Kr sensitivities and uncertainties}

Here we demonstrate in a detailed way how to use the sensitivity in order to calculate the impact of the nuclear uncertainties on the isotopic abundances. We focus on the sensitivity of $^{86}$Kr and $^{84}$Kr, which is affected by the branch point $^{85}$Kr, with 
a $\beta$-decay half-life of about 10~years \cite{RTR13,BBK00,ABG01}. 
The aim is to find all affecting global and local nuclear rates for the Kr isotopes and the impact of their uncertainties on the isotopic ratio, which can also be observed in presolar grains \cite{PGA06, LAA94}. 
Kr is of special interest since it can be measured in laboratories in presolar grains, condensed around old carbon-rich AGB stars before the formation of the solar system. From their analysis it is possible to measure isotopic abundances for $s$-process elements with high accuracy. 

After detecting all globally and locally affecting rates for the stable Kr isotopes during the TP and $^{13}$C-pocket (tables \ref{tab:Kr80}, \ref{tab:Kr82}, \ref{tab:Kr83}, \ref{tab:Kr84}, \ref{tab:Kr86}), we used these sensitivities to calculate uncertainties of the predicted Kr abundances resulting from uncertainties of the reaction rates, table~\ref{tab:uncer}. No sensitivities smaller than $\pm$0.1 in the $^{13}$C-pocket or the thermal pulse are listed.

\begin{table}[h]
\caption{Sensitivities for $^{80}$Kr\label{tab:Kr80}}
\begin{tabular}{r|ll}
 & $^{13}$C-pocket & TP \\
\hline
$^{79}$Se($\beta^\mathrm{-}$) 	                    & 0.828 & 0.83 \\
$^{22}$Ne($\alpha$,n)                               & $\mathrm{-}$ & 1.274 \\
$^{79}$Br(n,$\gamma$)                               & 0.37 & 0.421 \\
$^{74}$Ge(n,$\gamma$)                               & $\mathrm{-}$ & 0.745 \\
$^{72}$Ge(n,$\gamma$)                               & $\mathrm{-}$ & 0.457 \\
$^{78}$Se(n,$\gamma$)                               & $\mathrm{-}$ & 0.411 \\
$^{14}$N (n,p)                                      & 0.376 & $\mathrm{-}$ \\
$^{70}$Ge(n,$\gamma$)                               & $\mathrm{-}$ & 0.31 \\
$^{68}$Zn(n,$\gamma$)                               & $\mathrm{-}$ & 0.283 \\
$^{88}$Sr(n,$\gamma$)                               & 0.273 & $\mathrm{-}$ \\
$^{13}$C (p,$\gamma$)                               & 0.259 & $\mathrm{-}$ \\
$^{16}$O (n,$\gamma$)                               & 0.203 & $\mathrm{-}$ \\
$^{76}$Se(n,$\gamma$)                               & $\mathrm{-}$ & 0.188 \\
$^{69}$Ga(n,$\gamma$)                               & $\mathrm{-}$ & 0.172 \\
$^{73}$Ge(n,$\gamma$)                               & $\mathrm{-}$ & 0.158 \\
$^{71}$Ge(n,$\gamma$)                               & $\mathrm{-}$ & 0.125 \\
$^{90}$Zr(n,$\gamma$)                               & 0.108 & $\mathrm{-}$ \\
$^{22}$Ne(n,$\gamma$)                               & 0.191 & -0.148 \\
$^{24}$Mg(n,$\gamma$)                               & $\mathrm{-}$ & -0.104 \\
$^{64}$Ni(n,$\gamma$)                               & -0.182 & $\mathrm{-}$ \\
$^{58}$Fe(n,$\gamma$)                               & -0.217 & $\mathrm{-}$ \\
$^{12}$C (p,$\gamma$)                               & -0.286 & $\mathrm{-}$ \\
$^{25}$Mg(n,$\gamma$)                               & $\mathrm{-}$ & -0.375 \\
$^{13}$C ($\alpha$,n)                               & -0.404 & $\mathrm{-}$ \\
$^{56}$Fe(n,$\gamma$)                               & -0.198 & -0.214 \\
$^{80}$Kr(n,$\gamma$)                               & -0.548 & -1.021 \\
$^{79}$Se(n,$\gamma$)                               & -0.946 & -1.062 \\
\end{tabular}
\end{table}

\begin{table}[h]
\caption{Sensitivities for $^{82}$Kr\label{tab:Kr82}}
\begin{tabular}{r|ll}
 & $^{13}$C-pocket & TP \\
\hline
$^{22}$Ne($\alpha$,n)                               & $\mathrm{-}$ & 1.735 \\
$^{74}$Ge(n,$\gamma$)                               & $\mathrm{-}$ & 0.746 \\
$^{78}$Se(n,$\gamma$)                               & $\mathrm{-}$ & 0.59 \\
$^{80}$Se(n,$\gamma$)                               & $\mathrm{-}$ & 0.502 \\
$^{14}$N (n,p)                                      & 0.377 & $\mathrm{-}$ \\
$^{72}$Ge(n,$\gamma$)                               & $\mathrm{-}$ & 0.332 \\
$^{76}$Se(n,$\gamma$)                               & $\mathrm{-}$ & 0.269 \\
$^{88}$Sr(n,$\gamma$)                               & 0.258 & $\mathrm{-}$ \\
$^{13}$C (p,$\gamma$)                               & 0.248 & $\mathrm{-}$ \\
$^{79}$Se($\beta^\mathrm{-}$)     	            & $\mathrm{-}$ & 0.235 \\
$^{16}$O (n,$\gamma$)                               & 0.195 & $\mathrm{-}$ \\
$^{70}$Ge(n,$\gamma$)                               & $\mathrm{-}$ & 0.163 \\
$^{73}$Ge(n,$\gamma$)                               & $\mathrm{-}$ & 0.147 \\
$^{80}$Kr(n,$\gamma$)                               & $\mathrm{-}$ & 0.129 \\
$^{75}$As(n,$\gamma$)                               & $\mathrm{-}$ & 0.127 \\
$^{77}$Se(n,$\gamma$)                               & $\mathrm{-}$ & 0.109 \\
$^{90}$Zr(n,$\gamma$)                               & 0.103 & $\mathrm{-}$ \\
$^{22}$Ne(n,$\gamma$)                               & 0.184 & -0.203 \\
$^{57}$Fe(n,$\gamma$)                               & $\mathrm{-}$ & -0.112 \\
$^{64}$Ni(n,$\gamma$)                               & -0.131 & $\mathrm{-}$ \\
$^{24}$Mg(n,$\gamma$)                               & $\mathrm{-}$ & -0.142 \\
$^{79}$Se(n,$\gamma$)                               & $\mathrm{-}$ & -0.151 \\
$^{12}$C (p,$\gamma$)                               & -0.279 & $\mathrm{-}$ \\
$^{58}$Fe(n,$\gamma$)                               & -0.197 & -0.143 \\
$^{56}$Fe(n,$\gamma$)                               & -0.123 & -0.291 \\
$^{25}$Mg(n,$\gamma$)                               & $\mathrm{-}$ & -0.526 \\
$^{82}$Kr(n,$\gamma$)                               & -1.045 & -1.426 \\
\end{tabular}
\end{table}

\begin{table}[h]
\caption{Sensitivities for $^{83}$Kr\label{tab:Kr83}}
\begin{tabular}{r|ll}
 & $^{13}$C-pocket & TP \\
\hline
$^{22}$Ne($\alpha$,n)                               & $\mathrm{-}$ & 1.732 \\
$^{74}$Ge(n,$\gamma$)                               & $\mathrm{-}$ & 0.693 \\
$^{78}$Se(n,$\gamma$)                               & $\mathrm{-}$ & 0.606 \\
$^{80}$Se(n,$\gamma$)                               & $\mathrm{-}$ & 0.56 \\
$^{82}$Kr(n,$\gamma$)                               & $\mathrm{-}$ & 0.406 \\
$^{14}$N (n,p)                                      & 0.376 & $\mathrm{-}$ \\
$^{72}$Ge(n,$\gamma$)                               & $\mathrm{-}$ & 0.283 \\
$^{76}$Se(n,$\gamma$)                               & $\mathrm{-}$ & 0.273 \\
$^{88}$Sr(n,$\gamma$)                               & 0.257 & $\mathrm{-}$ \\
$^{13}$C (p,$\gamma$)                               & 0.247 & $\mathrm{-}$ \\
$^{16}$O (n,$\gamma$)                               & 0.195 & $\mathrm{-}$ \\
$^{79}$Se($\beta^\mathrm{-}$)              & $\mathrm{-}$ & 0.19 \\
$^{73}$Ge(n,$\gamma$)                               & $\mathrm{-}$ & 0.133 \\
$^{70}$Ge(n,$\gamma$)                               & $\mathrm{-}$ & 0.128 \\
$^{75}$As(n,$\gamma$)                               & $\mathrm{-}$ & 0.127 \\
$^{77}$Se(n,$\gamma$)                               & $\mathrm{-}$ & 0.112 \\
$^{81}$Br(n,$\gamma$)                               & $\mathrm{-}$ & 0.106 \\
$^{90}$Zr(n,$\gamma$)                               & 0.102 & $\mathrm{-}$ \\
$^{22}$Ne(n,$\gamma$)                               & 0.184 & -0.202 \\
$^{57}$Fe(n,$\gamma$)                               & $\mathrm{-}$ & -0.112 \\
$^{64}$Ni(n,$\gamma$)                               & -0.126 & $\mathrm{-}$ \\
$^{24}$Mg(n,$\gamma$)                               & $\mathrm{-}$ & -0.142 \\
$^{12}$C (p,$\gamma$)                               & -0.278 & $\mathrm{-}$ \\
$^{58}$Fe(n,$\gamma$)                               & -0.195 & -0.143 \\
$^{56}$Fe(n,$\gamma$)                               & -0.115 & -0.29 \\
$^{25}$Mg(n,$\gamma$)                               & $\mathrm{-}$ & -0.525 \\
$^{83}$Kr(n,$\gamma$)                               & -1.042 & -1.675 \\
\end{tabular}
\end{table}

\begin{table}[h]
\caption{Sensitivities for $^{84}$Kr\label{tab:Kr84}}
\begin{tabular}{r|ll}
 & $^{13}$C-pocket & TP \\
\hline
$^{22}$Ne($\alpha$,n)                               & $\mathrm{-}$ & 1.314 \\
$^{80}$Se(n,$\gamma$)                               & $\mathrm{-}$ & 0.548 \\
$^{78}$Se(n,$\gamma$)                               & $\mathrm{-}$ & 0.472 \\
$^{14}$N (n,p)                                      & 0.37 & $\mathrm{-}$ \\
$^{74}$Ge(n,$\gamma$)                               & $\mathrm{-}$ & 0.345 \\
$^{82}$Kr(n,$\gamma$)                               & $\mathrm{-}$ & 0.319 \\
$^{88}$Sr(n,$\gamma$)                               & 0.252 & $\mathrm{-}$ \\
$^{13}$C (p,$\gamma$)                               & 0.243 & $\mathrm{-}$ \\
$^{16}$O (n,$\gamma$)                               & 0.191 & $\mathrm{-}$ \\
$^{76}$Se(n,$\gamma$)                               & $\mathrm{-}$ & 0.185 \\
$^{81}$Br(n,$\gamma$)                               & $\mathrm{-}$ & 0.127 \\
$^{83}$Kr(n,$\gamma$)                               & $\mathrm{-}$ & 0.126 \\
$^{72}$Ge(n,$\gamma$)                               & $\mathrm{-}$ & 0.111 \\
$^{90}$Zr(n,$\gamma$)                               & 0.101 & $\mathrm{-}$ \\
$^{22}$Ne(n,$\gamma$)                               & 0.181 & -0.153 \\
$^{24}$Mg(n,$\gamma$)                               & $\mathrm{-}$ & -0.108 \\
$^{56}$Fe(n,$\gamma$)                               & $\mathrm{-}$ & -0.22 \\
$^{12}$C (p,$\gamma$)                               & -0.273 & $\mathrm{-}$ \\
$^{58}$Fe(n,$\gamma$)                               & -0.179 & -0.109 \\
$^{25}$Mg(n,$\gamma$)                               & $\mathrm{-}$ & -0.399 \\
$^{84}$Kr(n,$\gamma$)*                              & -0.428 & $\mathrm{-}$ \\
$^{84}$Kr(n,$\gamma$)                               & -0.612 & -0.607 \\
\end{tabular}
\end{table}

\begin{table}[h]
\caption{Sensitivities for $^{86}$Kr\label{tab:Kr86}}
\begin{tabular}{r|ll}
 & $^{13}$C-pocket & TP \\
\hline
$^{22}$Ne($\alpha$,n)                               & $\mathrm{-}$ & 2.515 \\
$^{84}$Kr(n,$\gamma$)                               & 0.417 & 1.408 \\
$^{85}$Kr(n,$\gamma$)                               & 0.946 & 0.84 \\
$^{82}$Kr(n,$\gamma$)                               & $\mathrm{-}$ & 0.386 \\
$^{80}$Se(n,$\gamma$)                               & $\mathrm{-}$ & 0.347 \\
$^{78}$Se(n,$\gamma$)                               & $\mathrm{-}$ & 0.203 \\
$^{83}$Kr(n,$\gamma$)                               & $\mathrm{-}$ & 0.174 \\
$^{13}$C ($\alpha$,n)                               & 0.144 & $\mathrm{-}$ \\
$^{81}$Br(n,$\gamma$)                               & $\mathrm{-}$ & 0.126 \\
$^{23}$Na(n,$\gamma$)                               & $\mathrm{-}$ & -0.117 \\
$^{32}$S (n,$\gamma$)                               & $\mathrm{-}$ & -0.122 \\
$^{57}$Fe(n,$\gamma$)                               & $\mathrm{-}$ & -0.163 \\
$^{58}$Fe(n,$\gamma$)                               & $\mathrm{-}$ & -0.201 \\
$^{24}$Mg(n,$\gamma$)                               & $\mathrm{-}$ & -0.206 \\
$^{56}$Fe(n,$\gamma$)                               & 0.133 & -0.421 \\
$^{22}$Ne(n,$\gamma$)                               & $\mathrm{-}$ & -0.292 \\
$^{84}$Kr(n,$\gamma$)*                              & -0.43 & $\mathrm{-}$ \\
$^{25}$Mg(n,$\gamma$)                               & $\mathrm{-}$ & -0.752 \\
$^{86}$Kr(n,$\gamma$)                               & -0.652 & -0.314 \\
$^{85}$Kr($\beta^\mathrm{-}$)              & -0.982 & -0.231 \\
\end{tabular}
\end{table}


With the obtained sensitivities for the Kr isotopes one can calculate
the uncertainties $\Delta N_j$ in the final abundance based on the recommended uncertainties
 of the rates $\Delta r_i$ with:
\begin{equation} 
\frac{\Delta N_j}{N_j} = \sqrt{\sum_i{ \left( s_{ij} \frac{\Delta r_i}{r_i}\right)^2 }}
\end{equation}
The largest contribution to the final uncertainty can be obtained with:
\begin{equation} 
\frac{\Delta N_j^{max}}{N_j^{max}} = \max_i{\left( s_{ij} \frac{\Delta r_i}{r_i}\right)} 
\end{equation}

The overall uncertainties are listed in tables~\ref{tab:uncer:13C} and \ref{tab:uncer:TP}. Note that despite significant experimental progress in determining neutron capture cross sections directly \cite{RBA04,RCH09,WCG15} or indirectly \cite{RTR13},
the by-far biggest contribution to the overall uncertainty comes from the neutron capture cross section on the unstable $^{85}$Kr. Current facilities are almost in the position to measure this cross section with sufficient accuracy \cite{CoR07}. Further developments are necessary to measure cross sections on isotopes with even shorter half-lives \cite{RHH04,ReL14}.

\begin{table}
\caption{Recommended uncertainties for rates with local effect on Kr \cite{DHK06,Sin02,Sie91,JKM01,AAR99,BBR88,HBH05,RTR13}. $^{71}$Ge(n,$\gamma$) was theoretically calculated based on \cite{RaT00} without error estimation.\label{tab:uncer}}
\begin{center}
\begin{tabular}{r|l|r|l}
Reaction & ${\Delta r}/{r}$ & Reaction & ${\Delta r}/{r}$ \\
\hline
$^{12}$C(p,$\gamma$)	 				& $\pm$10.1$\%$ & $^{13}$C(p,$\gamma$)	 				& $\pm$8.3$\%$\\
$^{13}$C($\alpha$,n) 					& $\pm$4.0$\%$ & $^{14}$N(n,p)	 					& $\pm$6.2$\%$\\
$^{16}$O(n,$\gamma$) 					& $\pm$10.5$\%$ & $^{22}$Ne($\alpha$,n) 					& $\pm$19.0$\%$\\
$^{22}$Ne(n,$\gamma$) 					& $\pm$6.9$\%$ & $^{23}$Na(n,$\gamma$) 					& $\pm$9.5$\%$\\
$^{24}$Mg(n,$\gamma$) 					& $\pm$12.1$\%$ & $^{25}$Mg(n,$\gamma$) 					& $\pm$6.3$\%$\\
$^{32}$S(n,$\gamma$) 					& $\pm$4.9$\%$ & $^{56}$Fe(n,$\gamma$) 					& $\pm$4.3$\%$\\
$^{57}$Fe(n,$\gamma$) 					& $\pm$10.0$\%$ & $^{58}$Fe(n,$\gamma$) 					& $\pm$5.2$\%$\\
$^{64}$Ni(n,$\gamma$) 					& $\pm$8.8$\%$ & $^{68}$Zn(n,$\gamma$) 					& $\pm$12.5$\%$\\
$^{69}$Ga(n,$\gamma$) 					& $\pm$4.3$\%$ & $^{70}$Ge(n,$\gamma$) 					& $\pm$5.7$\%$\\
$^{71}$Ge(n,$\gamma$) 					& $n.a.$ & $^{72}$Ge(n,$\gamma$) 					& $\pm$9.6$\%$\\
$^{73}$Ge(n,$\gamma$) 					& $\pm$19.3$\%$ & $^{74}$Ge(n,$\gamma$) 					& $\pm$10.4$\%$\\
$^{75}$As(n,$\gamma$) 					& $\pm$5.2$\%$ & $^{76}$Se(n,$\gamma$) 					& $\pm$4.9$\%$\\
$^{77}$Se(n,$\gamma$) 					& $\pm$17.0$\%$ & $^{78}$Se(n,$\gamma$)					& $\pm$16.0$\%$\\
$^{79}$Se(n,$\gamma$)					& $\pm$17.5$\%$ & $^{79}$Se($\beta^-$) 					& $\pm$12.9$\%$\\
$^{79}$Br(n,$\gamma$)					& $\pm$5.4$\%$ & $^{81}$Br(n,$\gamma$)					& $\pm$2.9$\%$\\
$^{80}$Se(n,$\gamma$) 					& $\pm$7.1$\%$ & $^{80}$Kr(n,$\gamma$)  					& $\pm$5.2$\%$\\
$^{82}$Kr(n,$\gamma$) 					& $\pm$6.7$\%$ & $^{83}$Kr(n,$\gamma$) 					& $\pm$6.2$\%$\\
$^{84}$Kr(n,$\gamma$)  					& $\pm$10.5$\%$ & $^{84}$Kr(n,$\gamma$)*			 		& $\pm$4.5$\%$\\
$^{85}$Kr(n,$\gamma$)					& $\pm$50$\%$ & $^{85}$Kr($\beta^-$)  					& $\pm$0.2$\%$\\
$^{86}$Kr(n,$\gamma$)					& $\pm$8.8$\%$ & $^{88}$Sr(n,$\gamma$)					& $\pm$1.8$\%$\\
$^{90}$Zr(n,$\gamma$)					& $\pm$4.7$\%$ & &\\
\end{tabular}
\end{center}
\end{table}


\begin{table}
\caption{Error estimation resulting from ${13}$C-pocket sensitivities and nuclear uncertainties for Kr isotopes
\label{tab:uncer:13C}}
\begin{center}
\begin{tabular}{r|l|l}
 Isotope & $\Delta N / N$ & $(\Delta N / N)^{max}$ \\
\hline
$^{80}$Kr                               & 20.6$\%$  & $^{79}$Se(n,$\gamma$) (16.3$\%$) \\
$^{82}$Kr                               & 8.7$\%$  & $^{82}$Kr(n,$\gamma$) (6.9$\%$) \\
$^{83}$Kr                               & 8.2$\%$  & $^{83}$Kr(n,$\gamma$) (6.3$\%$) \\
$^{84}$Kr                               & 8.4$\%$  & $^{84}$Kr(n,$\gamma$) (6.4$\%$) \\
$^{86}$Kr                               & 48.0$\%$  & $^{85}$Kr(n,$\gamma$) (47.3$\%$) \\
\end{tabular}
\end{center}
\end{table}

\begin{table}
\caption{Error estimation resuluting from TP sensitivities and nuclear uncertainties for Kr isotopes
\label{tab:uncer:TP}}
\begin{center}
\begin{tabular}{r|l|l}
 Isotope & $\Delta N / N$ & $(\Delta N / N)^{max}$ \\
\hline
$^{80}$Kr                               & 35.2$\%$  & $^{22}$Ne($\alpha$,n) (24.2$\%$) \\
$^{82}$Kr                               & 37.5$\%$  & $^{22}$Ne($\alpha$,n) (33.0$\%$) \\
$^{83}$Kr                               & 37.5$\%$  & $^{22}$Ne($\alpha$,n) (32.9$\%$) \\
$^{84}$Kr                               & 27.9$\%$  & $^{22}$Ne($\alpha$,n) (25.0$\%$) \\
$^{86}$Kr                               & 65.8$\%$  & $^{85}$Kr(n,$\gamma$) (42.0$\%$) \\
\end{tabular}
\end{center}
\end{table}

\section{Conclusions}

Because of the different conditions during the inter-pulse phase and the thermal pulse, only few rates have an impact in both conditions: $^{22}$Ne($\alpha$,n), $^{56}$Fe(n,$\gamma$), $^{58}$Fe(n,$\gamma$), $^{140}$Ce(n,$\gamma$), $^{142}$Nd(n,$\gamma$). 
Neutron poisons mostly affect the abundances produced in long-lived neutron-poor environments like the inter-pulse 
phase and are not important during short periods with higher neutron densities as in the convective thermal pulse. 

$^{14}$N is the strongest neutron poison in the $^{13}$C-pocket.
Competing neutron captures on the $s$ process path decrease the production of isotopes with large mass numbers on 
the chart of nuclides, as seen in the case $^{56}$Fe, $^{58}$Fe and $^{64}$Ni.
For the TP the most important rates, which affect the neutron density globally, are the neutron source $^{22}$Ne($\alpha$,n) and 
the neutron poison $^{25}$Mg.

An interactive graphical presentation of all data presented here is available at the URL: http://exp-astro.physik.uni-frankfurt.de/sensitivities/

\subsection*{Acknowledgments}
This work was 
partly supported by the HGF Young Investigators Project VH-NG-327, the Helmholtz International Center for FAIR, the Helmholtz Graduate School HGS-HIRe, the DAAD, the  DFG (SO907/2-1), the German Israeli Foundation,  NAVI, the EuroGenesis project MASCHE, and the European Research Council under the 
European Union's Seventh Framework Programme (FP/2007-2013)/ERC Grant 
Agreements n.$\:$615126. MP thanks for the support from the Ambizione grant of the SNSF (Switzerland) and acknowledges the NuGrid support. This work was also supported by the NSF grants PHY 02-16783 and PHY 09-22648 (Joint Institute for Nuclear Astrophysics, JINA) and the EU grant MIRG-CT-2006-046520. 

\bibliographystyle{elsarticle-num}
\bibliography{/home/reifarth/Texte/paper/refbib}

\end{document}